\documentclass[twocolumn]{aastex62}
\usepackage{graphicx}
\usepackage{amsmath,bm}
\newcommand{\Alfven}{Alfv\'en }
\newcommand{\Alfvenic}{Alfv\'enic }

\newcommand{\beunit}{nT$^{2}$}
\newcommand{\junit}{nA/m$^{2}$}
\newcommand{\vunit}{\mbox{km s$^{-1}$}}
\newcommand{\bunit}{\mbox{nT}}

\shorttitle{MMS Observation of Kinetic Signatures in the Alfv\'en Vortex}
\shortauthors{Wang et al.}

\begin{document}

\title{Magnetospheric Multiscale Observation of Kinetic Signatures in the Alfv\'en Vortex}


\author[0000-0003-3072-6139]{Tieyan Wang}
\affil{RAL Space, Rutherford Appleton Laboratory, Harwell Oxford, Didcot OX11 0QX, UK}

\author{Olga Alexandrova}
\affiliation{LESIA, Observatoire de Paris, Universit\'e PSL, CNRS, Sorbonne Universit\'e, Univ. Paris Diderot, Sorbonne Paris Cit\'e, 5 place Jules Janssen, 92195 Meudon, France}

\author{Denise Perrone}
\affiliation{Department of Physics, Imperial College London, London SW7 2AZ, UK}

\author{Malcolm Dunlop}
\affiliation{RAL Space, Rutherford Appleton Laboratory, Harwell Oxford, Didcot OX11 0QX, UK}
\affiliation{School of Space and Environment, Beihang University, Beijing 100191, China}

\author{Xiangcheng Dong}
\affiliation{School of Space and Environment, Beihang University, Beijing 100191, China}

\author{Robert Bingham}
\affiliation{RAL Space, Rutherford Appleton Laboratory, Harwell Oxford, Didcot OX11 0QX, UK}

\author{Yu. V. Khotyaintsev}
\affiliation{Swedish Institute of Space Physics, Uppsala, Sweden}

\author{C. T. Russell}
\affiliation{Institute of Geophysics and Planetary Physics, University of California, Los Angeles, CA, USA}

\author{B. L. Giles}
\affiliation{NASA Goddard Space Flight Center, Greenbelt, MD 20771}

\author{R. B. Torbert}
\affiliation{University of New Hampshire, Durham, New Hampshire, USA}

\author{R. E. Ergun}
\affiliation{Laboratory of Atmospheric and Space Physics, University of Colorado, Boulder, Colorado 80303, USA}

\author{J. L. Burch}
\affiliation{Southwest Research Institute San Antonio, San Antonio, TX 78238}

\begin{abstract}

Alfv\'en vortex is a multi-scale nonlinear structure which contributes to intermittency of turbulence. Despite previous explorations mostly on the spatial properties of the Alfv\'en vortex (i.e., scale, orientation, and motion), the plasma characteristics within the Alfv\'en vortex are unknown. Moreover, the connection between the plasma energization and the Alfv\'en vortex still remains unclear. Based on high resolution in-situ measurement from the Magnetospheric Multiscale (MMS) mission, we report for the first time, distinctive plasma features within an Alfv\'en vortex. This Alfv\'en vortex is identified to be two-dimensional ($k_{\bot} \gg k_{\|}$) quasi-monopole with a radius of ~10 proton gyroscales. Its magnetic fluctuations $\delta B_{\bot}$ are anti correlated with velocity fluctuations $\delta V_{\bot}$, thus the parallel current density $j_{\|}$ and flow vorticity $\omega_{\|}$ are anti-aligned. In different part of the vortex (i.e., edge, middle, center), the ion and electron temperatures are found to be quite different and they behave in the reverse trend: the ion temperature variations are correlated with $j_{\|}$, while the electron temperature variations are correlated with $\omega_{\|}$. Furthermore, the temperature anisotropies, together with the non-Maxwellian kinetic effects, exhibit strong enhancement at peaks of $|\omega_{\|}| (|j_{\|}|)$ within the vortex. Comparison between observations and numerical/theoretical results are made. In addition, the energy-conversion channels and the compressibility associated with the Alfv\'en vortex are discussed. These results may help to understand the link between coherent vortex structures and the kinetic processes, which determines how turbulence energy dissipate in the weakly-collisional space plasmas.

\end{abstract}

\keywords{plasmas --- turbulence}

\section{Introduction} \label{sec:intro}

Turbulence typically manifest itself as disordered, self-organized structures across different scales, resulting from nonlinear interactions of various physical properties with many degrees of freedom. Astrophysical plasmas such as interstellar medium and stellar atmosphere are found to exist in a turbulent state, and the solar-terrestrial environment is no exception \citep{1995SSRv...73....1T,1999PhPl....6.4137C,2010SSRv..156...89Z}. Understanding how the magnetized plasma turbulence works is challenging since it involves complex interactions between the electromagnetic fields and particles, leading to a state far from thermal equilibrium. Two fundamental properties have been found in explaining how the turbulence energy redistribute (usually referred to as ``cascade'') from large system scales to small kinetic scales. One is the power-law scaling of the turbulence energy spectrum at the so-called inertial range, where phenomenological approaches \citep[e.g.][]{1995tlan.book.....F,2003matu.book.....B} are used to describe the self-similar (scale invariant) fluctuations. However, the fully developed turbulence never displays pure scale invariance, instead it is characterized by bursty fluctuations emerging spontaneously. This property is referred to as intermittency, usually recognized by an increase of non-Gaussianity for the statistics of the fluctuations towards smaller scales (i.e., the scale dependent sharpening of the central peak and the heavy tail of the probability distribution function PDF in \citeauthor{2001P&SS...49.1193S} \citeyear{2001P&SS...49.1193S}). It is conjectured that, as the turbulence cascade proceed, the inhomogeneous transfer of energy toward small scales will lead to an uneven concentration of energy in limited volumes, thereby forming a series of patchy, phase correlated structures localized in space \citep{1995tlan.book.....F,2013SSRv..178..101A,2016JPlPh..82f5302C}. Although the physical mechanisms for the generation of coherent structures are still unclear, the intermittent events are believed to be crucial for the localized energy transfer and dissipation process.

Intermittency in hydrodynamics is depicted in high-amplitude tube-like vortex filaments \citep{1990Natur.344..226S}, where the vorticity is most likely aligned with the intermediate eigenvector of the strain rate \citep{1987PhFl...30.2343A}. These filament objects have two characteristic lengths. The larger one is comparable to the system size, at which the energy is injected. The smaller one is close to Kolmogorov dissipation length, where the molecular dissipation kicks in. Due to the inherently complex nature of the space plasma, the intermittent structures in this system exhibit more characteristic scales (e.g. magneto-fluid scale, ion and electron gyroscales) and topologies than that in the neutral fluid. Among various textures in plasma turbulence, current sheets, vortices, magnetic holes, solitons and shocks have been commonly observed \citep{1999AIPC..471..543V,2006JGRA..11112208A,2016ApJ...824...47L,2016JGRA..121.3870R,2016ApJ...823L..39G,
2016ApJ...826..196P,2017ApJ...849...49P,2017ApJ...836L..27H,2018ApJ...864...35Z}. Structures with strong gradient and relatively simple geometries (i.e., current sheet, rotational/tangential discontinuities) can be easily identified based on partial variance increment (PVI) techniques \citep{2008GeoRL..3519111G}. Subsequently, evidence for turbulence dissipation, plasma heating, and temperature anisotropy have been found near the current sheet/discontinuities in observations and simulations \citep{2007NatPh...3..236R,2012PhRvL.108z1103O,2012PhRvL.108z1102O,2009PhPl...16c2310P,2012PhRvL.108d5001S,2012PhRvL.109s5001W,
2013PhPl...20a2303K,2013ApJ...763L..30W,2013ApJ...772L..14W,2014EPJD...68..209P,2015ApJ...804L..43Z,
2015ApJ...804L...1C,2016NJPh...18l5001V,2016GeoRL..43.5969Y,2017JGRA..12211442V,2018PhRvL.120l5101C}. In particular, the location of plasma heating is reported to be in better agreement with places of vorticity than current density (i.e., simulation in \citeauthor{2016AIPC.1720d0003F} \citeyear{2016AIPC.1720d0003F} \citeauthor{2016ApJ...832...57P} \citeyear{2016ApJ...832...57P}), indicating the velocity gradient is as important as, if not more crucial than, the magnetic gradient.

In analogy to the hydrodynamic case, the non-linear coherent vortex structure also plays an important role in plasma dynamics and transport processes \citep{1978PhFl...21...87H,1985PhFl...28.1719S,1992swpa.book.....P,1994Chaos...4..227H}. These vortices tend to have a long lifetime and are widely observed in space, laboratory, and numerical simulation of plasma \citep{1988PhyS...38..841C,1990JGR....95.4333B,1996JGR...10113335V,2005Natur.436..825S,2000SSRv...92..423S,
2006JGRA..11112208A,2008NPGeo..15...95A,2008GeoRL..3515102A,2010NucFu..50d2002V,2015JPlPh..81a3207S}. An essential subset of these plasma vortices is known as \Alfven vortices, which can be viewed as cylindrical analogue of the non-linear \Alfven wave \citep{1992swpa.book.....P}. The \Alfven vortices have an axis nearly parallel to the unperturbed magnetic field, along which the shape is generally invariant. Thus, these vortices are quasi two-dimensional structures. The associated perpendicular magnetic fluctuations are linearly related with the perpendicular velocity fluctuations, but their relative amplitudes are not obligatory equal (as is the case in an \Alfven wave):
$\delta V_{\bot}/V_{A}=\xi\delta B_{\bot}/B_{0}$
where $\xi$ is not necessarily equal to 1. In addition, \Alfven vortices do not propagate along $\bm B_{0}$ in the plasma frame, hardly do they propagate in the perpendicular plane when the axis of the vortex is inclined with respect to $\bm B_{0}$, which are in contrast with \Alfven wave \citep{2012ApJ...746..147W}. After first being reported in the Earth's magnetosheath \citep{2006JGRA..11112208A,2008NPGeo..15...95A}, multi-scales quasi-bidimensional \Alfven vortices (with $k_{\bot} \gg k_{\|}$) have been identified in numerous space environments. For example, in slow solar wind \citep{2016JGRA..121.3870R,2016ApJ...826..196P}, in fast solar wind \citep{2016ApJ...824...47L,2017ApJ...849...49P}, and in Saturn's magnetosheath \citep{2008GeoRL..3515102A}. It seems that the intermittent structures in fast solar wind are dominated by \Alfven vortices \citep{2017ApJ...849...49P}, which agrees with the two-dimensional MHD turbulence model \citep{2017ApJ...835..147Z}.

Due to the limited temporal resolution of the plasma instrument on previous missions (i.e. the Cluster mission in \citet{2001AnGeo..19.1197E}), most of the \Alfven vortices have been identified solely on basis of magnetic field measurement \citep{2016JGRA..121.3870R,2016ApJ...826..196P,2017ApJ...849...49P}. An attempt to estimate velocity fluctuations $\delta \bm V$ using the electric and magnetic field data was done in \citep{2006JGRA..11112208A}. It was shown that indeed magnetic and velocity fluctuations are well aligned as expected for an \Alfven vortex. This result, however, needs to be confirmed by the direct measurements of the flow properties. Moreover, the knowledge of plasma characteristics, especially the velocity distribution functions accompanied with \Alfven vortices still remains blank. Although some results of electrons are discussed in \citet{2017ApJ...849...49P}, the 4 s resolution for the particles data was far from enough as compared to the vortex timescale. This led us to wonder, what is the detailed ion and electron behaviours within the \Alfven vortices? Is there any connection between \Alfven vortices and plasma kinetic effects? Thanks to the Magnetospheric Multiscale (MMS) mission \citep{2016SSRv..199....5B}, four closely-separated probes measures the turbulent magnetosheath region and provide high temporal resolution particle data during its burst mode (150 ms for ions and 30 ms for electrons). This allows us to overcome the major observational obstacles and resolve the sub-ion scales features of the vortex. In this paper, for the first time, (i) we verify the $\delta V - \delta B$ alignment using direct and independent measurements of velocity and magnetic fields and (ii) we report distinctive plasma kinetic signatures within the \Alfven vortex. The connection between coherent \Alfven vortex and plasma kinetic effects has thus been confirmed. Implications for the local energy conversion associated with the pressure strain interaction are discussed.

\section{Event Overview} \label{sec:data}
\begin{figure*}[ht!]
\figurenum{1}
\centering
\includegraphics [clip,trim=0cm 5cm 0cm 0cm, width=0.9\textwidth]{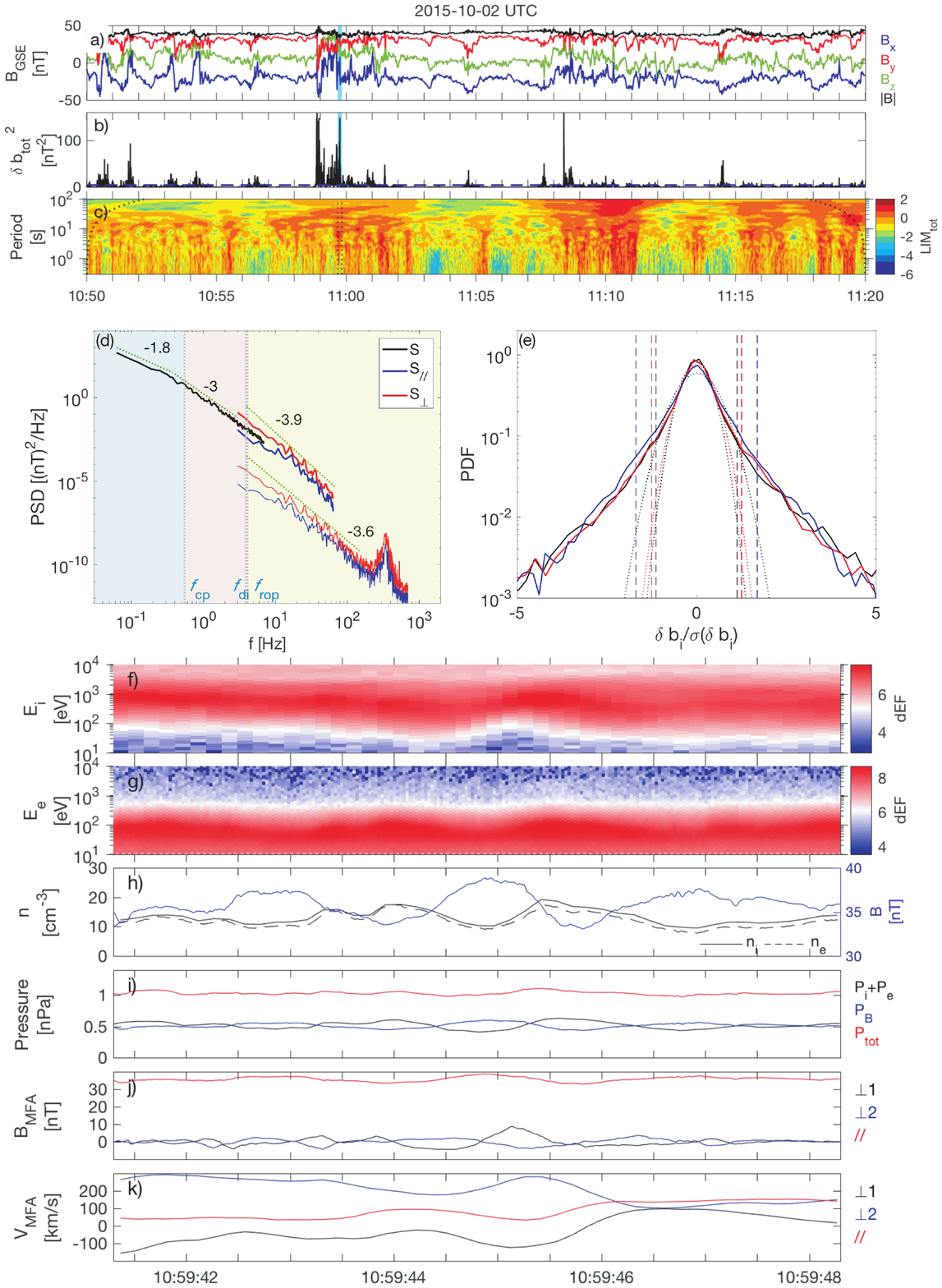}
\caption{
Overview of the event. a) Magnetic field in GSE coordinates and its magnitude during a 30 minutes interval on 2 October 2015. b) Total turbulent magnetic energy at ion scales (0.4 -- 3 Hz), where the horizontal blue dashed line indicates the threshold energy for the selection of intermittent intervals. The 7 s sub-interval marked by cyan shadow contains the structure of interest. c) logarithmic of $LIM_{total}$. d) PSD of the total magnetic field fluctuations during 10:50:00--11:20:00 UT (black curve), together with the PSDs (up to 700 Hz) of the parallel (blue) and perpendicular (red) magnetic field fluctuations during 10:59:41.5--10:59:48.5 UT. The spectra at 0.04 $<$ $f$ $<$ 60 Hz are calculated from FGM data, and the spectra at 3 $<$ $f$ $<$ 700 Hz are calculated from SCM data. Note that the original spectra at $f$ $>$ 3 Hz overlap with each other, thus the spectra based on SCM has been vertically shifted for better comparison. e) PDFs of the normalized ion-scale magnetic field fluctuations in the GSE coordinates, over plotted with the corresponding Gaussian distributions (dotted curves) and three standard deviations of each fit (vertical dashes). The closeup overview of the cyan shadowed interval include: f) Ion energy spectrogram. g) Electron energy spectrogram. h) Ion and electron density versus magnetic field strength. i) Magnetic pressure, plasma pressure, and total pressure. j) Magnetic field in MFA coordinates. k) Velocity field in the MFA coordinates.
\label{fig:fig1}}
\end{figure*}

We choose an interval during 10:50--11:20 UTC on 02 October, 2015, where MMS is located in the turbulent magnetosheath. The magnetic and electric field data are from the Flux Gate Magnetometer (FGM), Search Coil Magnetometer (SCM) and the Electric Double Probes (EDP) installed on the FIELDS suite, respectively \citep{2016SSRv..199..189R,2016SSRv..199..167E,2016SSRv..199..105T}. The three-dimensional particle data, in the form of ion and electron velocity distribution functions (VDFs) and the associated plasma moments (i.e., density, velocity, temperature, and pressure), are from the Fast Plasma Investigation (FPI) \citep{2016SSRv..199..331P}. In the GSE coordinates, the magnetic field is generally stable around $\bm B_{0}$ = 39.2 (--0.56, 0.82, 0.13) {~\bunit} throughout the interval as shown in Figure \ref{fig:fig1}a, while a few discontinuities with increase of $B_{z}$ and decrease of $B_{x}$ and $B_{y}$ can be spotted. The mean flow speed is around $\bm V_{0}$ = 170 (--0.7, 0.7, --0.1){~\vunit} and it has an angle of about 35$^\circ$ with respect to $\bm B_{0}$. The total magnetic field fluctuations energy, which is used to select the interval of interest, exhibit large variations as plotted in Figure \ref{fig:fig1}b. The sub interval marked by cyan colour contains the structure of interest and it will be further analysed in more details. The relevant plasma parameters during these two intervals are listed in Table \ref{tab:table}. To quantify the turbulence energy across different scales, Figure \ref{fig:fig1}d presents the trace of the power spectrum density (PSD) of the magnetic field. The black curve presents the results for the total magnetic field during the 30-minute interval. It can be seen that the PSD follows a power law $f^{-1.8}$ at [0.06, 0.38] Hz, suggesting a fluid-like behaviour in the inertial range. Then it steepens and follows $f^{-3}$ between [0.4, 3] Hz, where the spectral break generally matches with the proton cyclotron frequency $f_{cp}$ and the Doppler-shifted proton inertial length $f_{di}$, under the assumption of wave vector parallel to the plasma flow. Notice that the perpendicular ion plasma beta $\beta_{\bot i}=(\rho_{i}/d_{i})^2$ is of the order of one in our event, thus $f_{di}$ and the Doppler-shifted proton gyroradius $f_{\rho p}$ cannot be separated easily. Finally, the PSD steepens again and follows $f^{-3.9}$ between [4, 70] Hz.

\begin{deluxetable}{lccBccBcc}
\tablecaption{Plasma parameters during the intervals of interest \label{tab:table}}
\tablehead{
\colhead{} & \colhead{Interval 1 (UT)} & \colhead{Interval 2 (UT)} \\
\colhead{} & \colhead{10:50:00--11:20:00} & \colhead{10:59:41.5--10:59:48.5}
}
\startdata
$\bm B$ (nT) & 39 (--0.56, 0.82, 0.13) & 36 (--0.34, 0.81, 0.47) \\
$\bm V$ (\vunit) & 170 (--0.7, 0.7, --0.1) & 242 (--0.7, 0.5, --0.5) \\
$\theta_{\bm {BV}}$ (deg) & 35 & 67 \\
$N$ (cm$^{-3}$) & 14 & 13 \\
$T_i$ (eV) & 198 & 216 \\
$T_e$ (eV) & 31 & 31 \\
$T_{i,\|}/T_{i,\bot}$ & 0.8 & 1.0 \\
$T_{e,\|}/T_{e,\bot}$ & 1.4 & 1.7 \\
$\beta_i$ & 0.7 & 0.9 \\
$\beta_e$ & 0.1 & 0.1 \\
$\rho_p$ (km) & 55 &  60 \\
$d_i$ (km) & 60 &  62 \\
$f_{cp}$, $f_{\rho p}$, $f_{di}$ (Hz) & 0.59, 2.9, 2.7 &  0.55, 4.0, 3.9 \\
\enddata
\end{deluxetable}

For the transition range around ion scales, the turbulent energy is believed to further cascade and the kinetic physics begins \citep{2013SSRv..178..101A,2015rsta..11...11K}. Various forms of coherent structures (i.e., current sheet, vortex filaments) are usually found to reside near this range (i.e. between the end of the MHD range and proton scales in \citeauthor{2016ApJ...826..196P} \citeyear{2016ApJ...826..196P}; \citeauthor{2017ApJ...849...49P} \citeyear{2017ApJ...849...49P};
\citeauthor{2016ApJ...824...47L} \citeyear{2016ApJ...824...47L}). Hence in the following analysis we focus on fluctuations of similar scales, where the frequency ranges is chosen as [0.4, 3] Hz, and the timescale is [0.3, 3] s. It has been shown recently that, in case of a collisionless turbulent system as the solar wind, the intermittency, non-Gaussian fluctuations, and phase coherence of magnetic field components are interrelated \citep{2017ApJ...849...49P}. We expect this relation to be present here and thus take similar procedures as in \citet{2017ApJ...849...49P} to search for the intermittent events. The first step is to reconstruct the fluctuations using the band-pass filter based on wavelet transforms \citep{1998BAMS...79...61T,2012ApJ...745L...8H,2014JGRA..119.9527W,2016ApJ...826..196P}. The magnetic field fluctuations are thus defined as,
\begin{equation}
\label{equ:equ1}
    \delta b_{i}(t)=\frac{\delta j \delta t^{1/2}}{C_{\delta}\psi_0(0)}\sum\limits_{j=j1}^{j2}{\frac{\widehat{\mathcal W}_i(\tau_j),t)}{\tau_j^{1/2}}},
\end{equation}
Where $i$ represent the magnetic field components, $j$ represent the scale index, $\delta j$ is the constant scales step, $\widehat{\mathcal W}_i$ is the real part of the wavelet coefficient $\mathcal W_i$. $C_{\delta}$ = 0.776. $\psi_0$ is the Morlet mother function and $\psi_0(0)=\pi^{-1/4}$ at time $t$=0 \citep{1998BAMS...79...61T}. $\tau_{j1}$ and $\tau_{j2}$ are taken to be 0.3 and 3 s, respectively. The second step is to determine the threshold energy as defined by
$\varepsilon_T=\sum\limits_{i=1}^{3}(3\sigma_G(\delta b_i))^2$,
where $\sigma_G(\delta b_i)$ is the standard deviations of the Gaussian fit for each magnetic field fluctuations components. $\sigma_G(\delta b_i)$ are fitted to be 0.36, 0.32, 0.49 \bunit, leading to a $\varepsilon_T$ $\sim$ {~4.3 \beunit}. From a statistical point of view, 99.7 \% of all the values in Gaussian distribution are within $3\sigma_G(\delta b_i)$ from the mean. In other words, the events whose total energy $\delta b_{tot}^2=\sum\limits_{i=1}^{3}(\delta b_i)^2$ are larger than $\varepsilon_T$ could contribute to the non-Gaussian part of the distributions. Figure \ref{fig:fig1}e presents the PDFs of the normalized magnetic field fluctuations $\delta b_i/\sigma(\delta b_i)$ together with their Gaussian fits. The presence of clear non-Gaussian tails suggest the abundance of intermittent events during the whole interval. The last step is to locate these events. As seen in Figure \ref{fig:fig1}b, there are approximately 14 events (with at least 10 s in duration) with $\delta b_{tot}^2$ larger than 5 \beunit. We have picked one interval during 10:59:41.5-10:59:48.5 UT for further studies. This event is characterized by large magnetic energy $\sim$ 150 {~\beunit} and strong flow vorticity up to $\sim$ 1.4 /s as compared to the mean value of $\sim$ 0.4 /s.  Furthermore, the local intermittency measure ($LIM$) exhibit an extension of temporal scale from several tens of seconds to sub seconds, as seen in the $LIM_{total}$ of Figure \ref{fig:fig1}c.  The $LIM$ spectrogram, as a function of time and scales, is computed as the instantaneous energy of fluctuations normalized to its mean value over the studied time interval
 \citep{1992AnRFM..24..395F}:
\begin{equation}
\label{equ:equ2}
    LIM_{\|,\bot,total}=|\mathcal W_{\|,\bot,total}|^2/ \left \langle| \mathcal W_{\|,\bot,total}|^2 \right \rangle_t
\end{equation}
Where $|\mathcal W_{total}|^2=|\mathcal W_{\|}|^2+|\mathcal W_{\bot}|^2$ is the total fluctuation energy. For the coherent structures (space or time localised energetic events), one of its intrinsic properties is the energy coupling over many scales \citep{1992AnRFM..24..395F,1995tlan.book.....F}, which is usually manifested in the spanning of $LIM$ at a wide range of spatial scales (or temporal scales if the Taylor frozen-in hypothesis is assumed) \citep{2016ApJ...824...47L,2017ApJ...849...49P}. While for the wave phenomenon, the energy distribution is typically focused around a certain frequency, i.e. \Alfven ion {cyclotron} \citep{2004JGRA..109.5207A} or electron cyclotron waves \citep{2014ApJ...796....5L}. Therefore, the extension of $LIM$ provide evidence of coupling from MHD to sub ion scales and hence imply the presence of a coherent structure \citep{1992AnRFM..24..395F,1995tlan.book.....F,2016ApJ...824...47L,2017ApJ...849...49P}. Note that since $LIM_{\|}$, $LIM_{\bot}$, and $LIM_{total}$ exhibit nearly the same features in this event, only $LIM_{total}$ is presented here.

The overview of this 7 s event is presented in Figure \ref{fig:fig1}f -- \ref{fig:fig1}k, together with the PSDs of the perpendicular and parallel magnetic field fluctuations (denoted as $S_{\bot}$ and $S_{\|}$) displayed in Figure \ref{fig:fig1}d. The ion and electron differential energy spectrograms exhibit fluctuations in their energy levels and magnitudes (Figure\ref{fig:fig1}f and Figure\ref{fig:fig1}g), in correspondence with the density variations (Figure\ref{fig:fig1}h). However, the total pressure is almost stable, where the plasma pressure is balanced by the magnetic pressure (Figure\ref{fig:fig1}i). Interestingly, large amplitude magnetic field fluctuations ($>$10 \bunit) and velocity fluctuations ($>$50 \vunit) are found to be localized in time (within 6 seconds). In addition, these fluctuations are dominant in the perpendicular direction as seen in Figure \ref{fig:fig1}j and \ref{fig:fig1}k ($\delta B_{\bot}/B_0$ $\sim$ 0.37, $\delta B_{\|}/B_0$ $\sim$ 0.1, $\delta V_{\bot}/V_A$ $\sim$ 0.38, $\delta V_{\|}/V_0$ $\sim$ 0.1). The time series are presented in the mean field-aligned (MFA) system, where the $\bm z$ direction corresponds to a 3 s running averaged of the magnetic field in the GSE coordinates, the $\bm y$ direction is obtained from the cross product of the $\bm z$ vector and the spacecraft location in the GSE coordinates, and the $\bm x$ direction is the cross product of $\bm y$ and $\bm z$ directions. Moreover, the slope for the perpendicular fluctuations are close to --4 (i.e., --3.6 for the result based on FGM and --3.9 for the result based on SCM). These features give some hint to the presence of the incompressible \Alfven vortex structure, which has localized in space strong perpendicular magnetic field fluctuations and a theoretical --4 slope for the PSD as due to the discontinuity of the parallel current density at the vortex boundary \citep{2008NPGeo..15...95A}. Notice that the ``bump'' in the PSD near the electron cyclotron frequency corresponds to parallel whistler emissions within the structure, which is out of the scope of the present paper but it will be studied in a future work.

\section{Kinetic signatures in the Alfv\'en vortex} \label{sec:results}

\subsection{Identification of the Alfv\'en vortex}
\label{subsec:topology}

\begin{figure*}[ht!]     
\figurenum{2}
\includegraphics [clip,trim=0cm 15cm 0cm 0cm, width=1.00\textwidth]{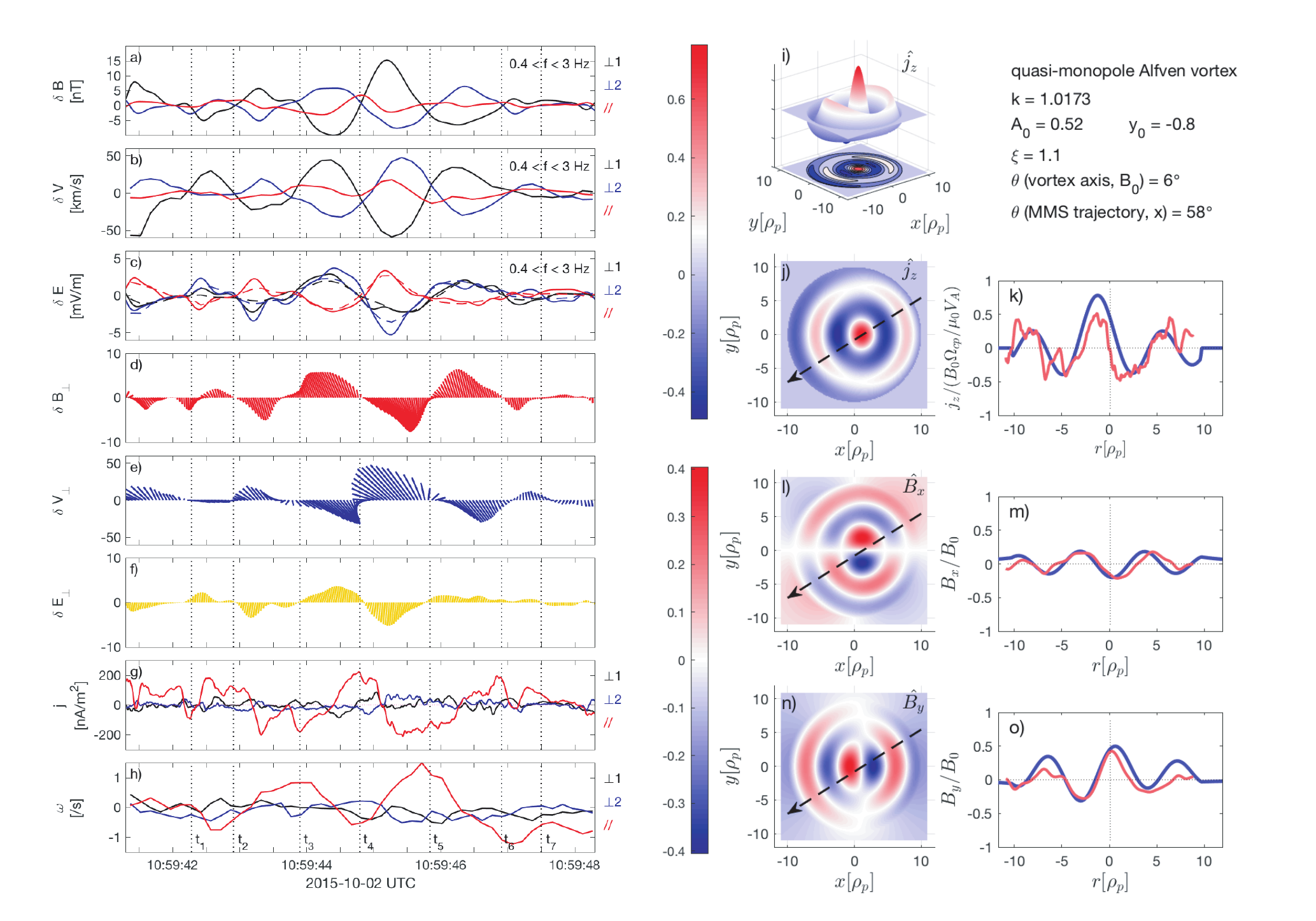}
\caption{Identification of the \Alfven vortex. a) Magnetic field fluctuations in MFA coordinates. b) Same as a), here for the velocity field. c) Same as a), here for the electric field. The solid lines represent direct $\delta E$ measurements while the dashed lines correspond to $-\delta (V_i \times B)$ d) Hodogram demonstrations of perpendicular magnetic field. e) Same as d), here for the velocity field. f) Same as d), here for the electric field. g) current density in the MFA coordinates. h) Flow vorticity in the MFA coordinates calculated from ion velocity. Seven times (from $t_1$ to $t_7$) within the vortex are marked as vertical dashed lines. i) 3D representation of the current density pattern based on dipolar \Alfven vortex model, with its parameters listed on the right side. j) parallel current pattern overplotted with the spacecraft trajectory. k) Comparison between the vortex solution with the MMS observation for the parallel current. Note that the current is normalized by $B_0 \Omega_{cp}/\mu_0V_A$, the magnetic field is normalized by mean magnetic field $B_0$, the spatial lengths are normalized by proton gyroradius $\rho_p$, and the time series from MMS have been transferred to spatial series based on the timing results of around 200 \vunit. l) Same as j) but for $B_{\bot 1}$. m) Same as k but the $B_{\bot 1}$. n) Same as j) but for $B_{\bot 2}$. o) Same as k but for $B_{\bot 2}$.
\label{fig:fig2}}
\end{figure*}

To better determine the nature/type of this structure, we present more detailed analysis of the small-scale electromagnetic and velocity field as well as current density, flow vorticity in Figure \ref{fig:fig2}a--\ref{fig:fig2}h. As seen in Figure \ref{fig:fig2}a--\ref{fig:fig2}c, during the interval which start from $t_1$ and terminate at $t_7$, the perpendicular components of $\delta \bm B$, $\delta \bm V$, and $\delta \bm E$ exhibit 5 polarity reversals marked as $t_2$, $t_3$, $...$, $t_6$. The variations of the field direction, also revealed in feather plots of the $\delta B_{\bot}$, $\delta V_{\bot}$, $\delta E_{\bot}$ (Figure \ref{fig:fig2}d--\ref{fig:fig2}f), are reminiscent of vortices. The directional changes of $\delta B_{\bot}$ and $\delta V_{\bot}$ correspond to the extrema of the parallel current density $j_{\|}$ and flow vorticity $\omega_{\|}$, which are much larger than $j_{\bot}$ and $\omega_{\bot}$ (Figure \ref{fig:fig2}g and \ref{fig:fig2}h). Here the current and vorticity are calculated by applying the curlometer method to the magnetic and ion velocity field, respectively \citep{2002JGRA..107.1384D}, where the validity of the methods has been verified in recent MMS observations of ion scale currents \citep{2018JGRA..123.5464D}. In addition to the perpendicular field reversals, we observe clear anti-correlation of $\delta B_{\bot}$ and $\delta V_{\bot}$, satisfying the relation $\delta V_{\bot}/V_A = -\xi B_{\bot}/B_{0}$, where $\xi_{\bot1}$ is $\sim$ 1.14 and $\xi_{\bot2}$ is $\sim$ 1.03 for the two perpendicular directions. The small difference between the two coefficients may be related to linear regression error, which is $\sim$ 0.1 in this case. The unique $\delta B_{\bot}$, $\delta V_{\bot}$, $j_{\|}$, and $\omega_{\|}$ features indicate again the presence of coherent \Alfvenic vortex, which manifest itself as a two-dimensional tube-like structure with quasi field-aligned current \citep{2006JGRA..11112208A}. In addition, the direct $\delta \bm E$ observation (solid lines in Figure \ref{fig:fig2}c) nearly match with $\delta \bm E=-\delta(\bm V_i \times \bm B)$ (dash lines in Figure \ref{fig:fig2}c), which verifies the assumption that the electric field at scales of the Alfven vortex can be approximated by the convection term of the Ohm's law \citep{2006JGRA..11112208A}.

To provide more evidence for the existence of \Alfven vortex, we first obtain the orientation and motion of the structure from four spacecraft measurements, and then compare the observations with \Alfven vortex solutions to determine more parameters of the vortex (e.g. type, inclination, radius). Here the timing method \citep{1998ISSIR...1..249S,2006JGRA..11112208A} is used to calculate the normal direction $n$ and the propagation velocity $V_n$, where the accuracy is guaranteed by the clear time-shift ($\sim$ 0.1 s) of the signals measured by four spacecrafts with separations $\sim$ 20 km. The inferred angle between $\bm n$ and the local $\bm B_0$ is around 86.8$^{\circ}$. This result is in contrast with the minimum variance analysis (MVA) from single spacecraft measurement, which gives a normal (or wave vector) direction nearly parallel to $\bm B_0$ ($\theta_{\bm k,\bm B_0}$ $\sim$ 8 $\pm$ 5$^{\circ}$). Indeed, the difference of the ``normal'' directions from different methods favors the presence of a localized cylindrical vortex rather than a parallel propagating plane wave. For this tube-like topology, its axis is given by the minimum variance direction (i.e. along $\bm B_0$), while the normal of its surface is given by the timing results (i.e. perpendicular to $B_0$). Besides that, the propagation velocity $\bm V_n$ is $\sim$ (70, 189, 18) $\pm$ 20 \vunit in the MFA frame, and the perpendicular velocity $\bm V_\bot n$ is $\sim$ (70, 190, 0) $\pm$ 20 \vunit, being slightly larger than the perpendicular flow speed (25, 190, 0) \vunit. Hence the vortex barely propagates in the plasma rest frame (with the perpendicular propagation speed $u_\bot$ $\sim$ 45 $\pm$ 20 \vunit, $u_\bot/V_A$ $\sim$ 0.18$\pm$0.08).

Among various models describing the localized \Alfven vortex filaments, one simple case is the specific nonlinear solutions of the ideal incompressible MHD system \citep{1992swpa.book.....P,2008NPGeo..15...95A,2017arXiv170502913J}, which satisfies the generalized \Alfven relation
\begin{equation}
\label{equ:equ3}
    \psi \propto \xi A
\end{equation}

Here $\psi$ is the flux function, which relate to the transverse velocity fluctuations $\delta \bm V_{\bot}=z \times \nabla \psi$, $A$ is the magnetic potential, which relate to the transverse magnetic fluctuations $\delta \bm B_{\bot}=\nabla A \times \bm z$, and $\bm z$ is the magnetic field direction.

The \Alfven vortex solution in the vortex plane $(x,\eta)$ reads
\begin{equation}
\label{equ:equ4}
\left\{\begin{matrix}
A=A_0(J_0(kr)-J_0(ka))+\frac{u}{\xi} \frac{x}{kr}(kr-2\frac{J_0(kr)}{J_1(ka)}),& r<a\\
A=a^2\frac{u}{\xi}\frac{x}{r^2},& r \geq a
\end{matrix}\right.
\end{equation}

The analytical expression depends on the axial distance to the vortex center $r=\sqrt{x^2+\eta^2}$, where $\eta$ is defined as $\eta=y+uz/\xi-ut$, with $u$ being the vortex propagation speed in the vortex plane $(x,\eta)$. If the angle between the vortex axis and background magnetic field $\bm B_0$ is $\theta_{vortex}$, then $\xi$ is defined as $\xi=u/\tan(\theta_{vortex})$. Inside the vortex core ($r<a$), the first term, in the form of the Bessel function of zeroth order $J_0$, describes the monopole component with an arbitrary amplitude $A_0$. The second term, in the form of Bessel function of the first order $J_1$, describes the dipolar components relating to the vortex inclination/propagation effects. The amplitude of the dipolar component depends on $u/{\xi}$ that is $\tan(\theta_{vortex})$. Outside the vortex core ($r>a$), only dipole component is non-zero and it decays at infinity as a power-law $\sim$ $1/r$. Note that the continuity of the solution at $r=a$ requires $J_1(ka)=0$. In the limit of $u=0$, the solution is axial symmetric and the vortex is a strictly field-aligned monopole. In the limit of $A_0=0$, the solution is axial asymmetric and the vortex is a strictly dipole. Under other circumstances, the combined solution of the two terms in equation (\ref{equ:equ4}) depicts a mixed scenario (e.g., monopole sitting on a dipole).

We choose the quasi-monopole model to fit the data since the observed time series of $j_{\|}$ appears to be symmetric around the half time of the interval, resembling the monopole configuration. In addition, we also consider the propagation effect by choosing the nonzero angle between the vortex axis and $\bm B_0$, which is different from the strictly aligned case for the monopole. In particular, the vortex diameter is estimated to be around 20 $\rho_p$ to match the timing results of $\sim$ 1200 km. The third zero of the Bessel function is selected ($ka$ = 10.17) to approximate the triple peaks of the current density. The angle between the vortex axis and $\bm B_0$ is chosen as $\theta_{vortex} \sim 6^{\circ}$, nearly in agreement with MVA analysis. We note that the perpendicular speed of the vortex $u_{\bot}=\tan(\theta_{vortex})V_A$ $\simeq$ 0.13 $\pm$ 0.09 $V_A$ as well as the angle of the MMS trajectory $\theta_{MMS}$ $\sim$ 58$^{\circ}$, are in qualitative agreement with the timing results of 0.18 $\pm$ 0.08 $V_A$, and 69$^{\circ}$, respectively, suggesting the self-consistency of the fitting process. The modeled current and magnetic field results are shown in Figure \ref{fig:fig2}i--\ref{fig:fig2}n, with the dashed lines representing the virtual trajectory of MMS, and the parameters listed on the right-hand side of Figure \ref{fig:fig2}i. It can be seen from Figure \ref{fig:fig2}i and \ref{fig:fig2}j that the current has three positive peaks and two negative peaks located at the edge ($r \sim 10$ $\rho_{p}$), center ($r \sim 0$ $\rho_{p}$) and middle part ($r \sim 3$ $\rho_{p}$) of the vortex. As a result, three layers of the azimuthal magnetic field are visible within the vortex, which agree with the $B_x$, $B_y$ contours in Figure \ref{fig:fig2}l and \ref{fig:fig2}n. The pattern for the quasi-monopole solution here closely resembles the circular symmetric monopolar solution, but for a slight asymmetry along the $x=0$ axis. This is attributed to the major influence from the first term (symmetric) in the vector potential (see equation (\ref{equ:equ4})) than the minor, asymmetric effect from the second term. Figure \ref{fig:fig2}k, \ref{fig:fig2}m, \ref{fig:fig2}o present the direct comparison between the modeled results and the observation. The agreement between the two results confirms the feasibility of the quasi-monopole description for the observed \Alfven vortex.

\subsection{Kinetic signatures within the Alfv\'en vortex}
\label{subsec:kinetic}
\begin{figure*}
\figurenum{3}
\centering
\includegraphics [clip,trim=0cm 1cm 0cm 0cm, width=0.86\textwidth]{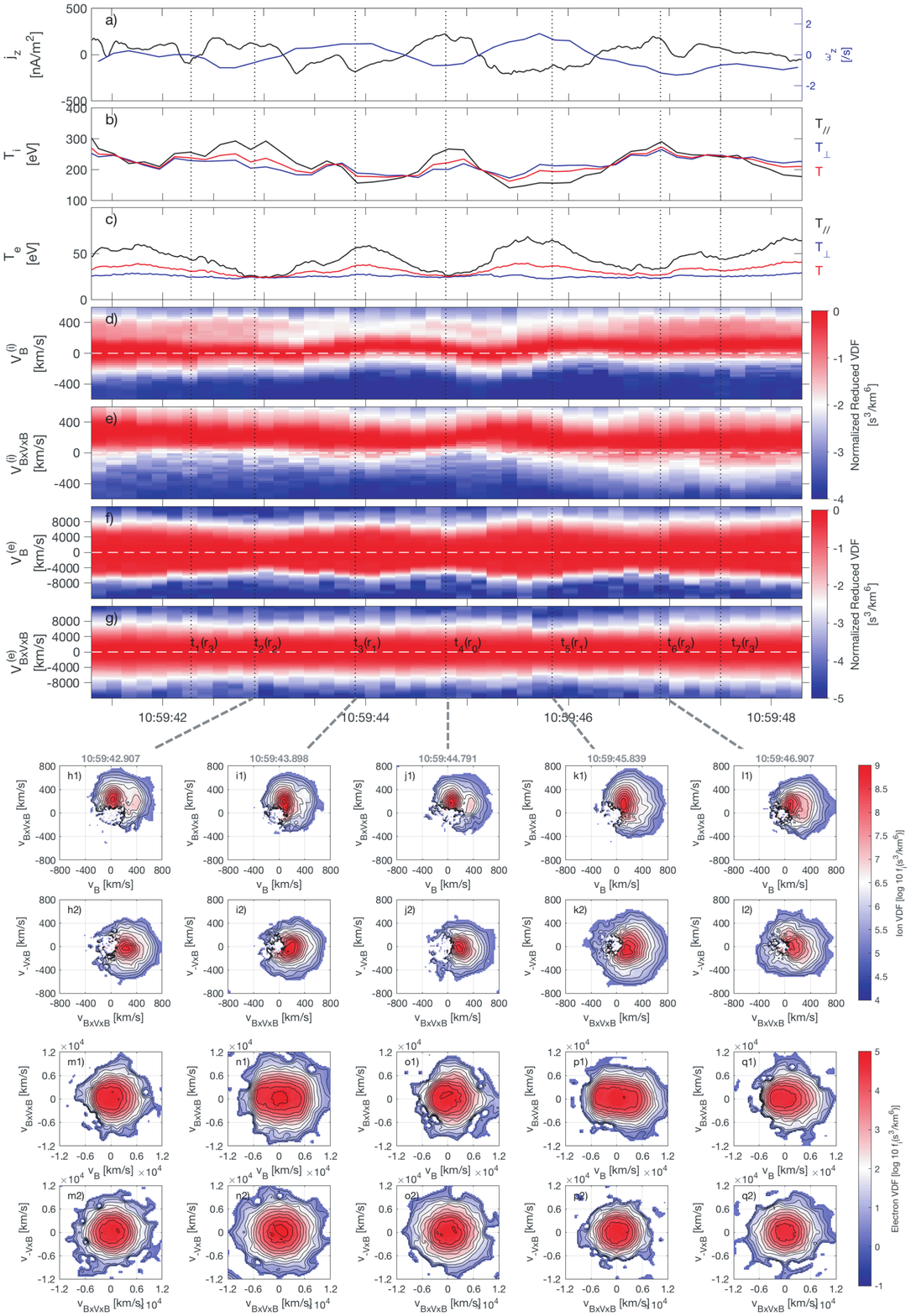}
\caption{Kinetic signatures observed in the \Alfven vortex. a) Current density versus flow vorticity in the MFA coordinates. b) Ion parallel, perpendicular and total temperature. c) Same as c) but for electrons. d) Normalized reduced ion velocity distribution functions (NR-iVDFs) in the parallel direction. e) Same as d) but in the perpendicular direction. f) Normalized reduced electron velocity distribution functions (NR-eVDFs) in the parallel direction. g) Same as f) but in the perpendicular direction. h1) -- l1) Projection of the iVDFs in the local ($\bm B$, $\bm B \times \bm V \times \bm B$) plane, taken during the times marked by the vertical dotted lines. h2) - l2) Projection of the iVDFs in the local ($\bm B \times \bm V \times \bm B$, $-\bm V \times \bm B$) plane.  m1) - q1) Same as h1) - l1) but for electrons. m2) - q2) Same as h2) - l2) but for electrons. Note that the projection of VDFs on planes constructed by $\bm B_0$, $\bm V_0$ exhibit similar features as compared to the results shown here.
\label{fig:fig3}}
\end{figure*}

Now we have gained the geometrical properties of the \Alfven vortex, it is thus possible to study the plasma features at different locations within it. To underline the specific observation made at certain location, Figure \ref{fig:fig3} begins with the same current density and vorticity as in Figure \ref{fig:fig2}, and then present the temperatures as well as the VDFs for ions and electrons, respectively. In addition, the vertical dashed lines are kept the same as Figure \ref{fig:fig2} so as to mark the different times and locations within the vortex, where $t_1$ and $t_7$ correspond to outer edge ($r_3$), $t_2$ and $t_6$ correspond to the inner edge ($r_2$), $t_3$ and $t_5$ correspond to the middle ($r_1$), and $t_4$ correspond to the center ($r_0$) of the vortex.

Three distinct temperature features in association with the strong magnetic/velocity field gradient can be identified: (1) There is a correlation between the total ion/electron temperature and parallel current density $j_{\|}$/vorticity $\omega_{\|}$. Ions are relatively hotter ($\sim$ 260 eV) near $r_2$ and $r_0$, when $j_{\|}$ reaches its local maxima. Yet their temperatures are colder ($\sim$ 180 eV) near $r_1$, when $j_{\|}$ is at its local minima. In contrast, electrons are relatively colder ($\sim$ 25 eV) at $j_{\|, max}$($\omega_{\|, min}$), and are hotter ($\sim$ 60 eV) at $j_{\|, min}$($\omega_{\|, max}$). (2) The temperature anisotropy $T_{\|}/T_{\bot}$, is correlated with $j_{\|}$($\omega_{\|}$) (see Figure \ref{fig:fig4}b and \ref{fig:fig4}c). At locations near $r_2$ and $r_0$, $T_{\|}$ is larger than $T_{\bot}$ for ions, while $T_{\|}$ and $T_{\bot}$ are almost the same for electrons. At locations near $r_1$, $T_{\|}$ is smaller than $T_{\bot}$ for ions, while $T_{\|}$ is larger than $T_{\bot}$ for electrons. (3) The electron temperatures behave in an opposite trend as compared with ions, which may reflect a balanced energy allocation between these two components. In addition, the redistribution of energy mainly happens in the parallel direction.

Ion energization and anisotropization has been revealed to occur near, but not centered on current structures in recent two-dimensional hybrid simulations and theories \citep{2016AIPC.1720d0003F,2016ApJ...832...57P,2016NJPh...18l5001V,
2016PhRvE..93e3203D}. To our knowledge, the MMS observations reported here provide the first evidence of plasma temperature anisotropy inside the vortex structure. The ions' behaviors, in particular, verify the correlation between temperature anisotropy and the out-of-plane vorticity observed in \citep{2016AIPC.1720d0003F,2016ApJ...832...57P,2016NJPh...18l5001V}. In these studies, due to the different spatial distribution of the vortex and the current sheets (i.e. the vorticity is less filamentary then the current sheets and sometimes eddies are formed on the flank of the planar current sheets), the ion temperature anisotropy displays different correlation with $|\omega_{\|}|$ than with $|j_{\|}|$ \citep{2016AIPC.1720d0003F,2016ApJ...832...57P}. For the vortex of \Alfvenic nature reported here though, we have explored another scenario where the anti-phased perpendicular magnetic and velocity field implies the alignment of vorticity and current density, hence the correlation between ion temperature anisotropy with $|\omega_{\|}|$ and $|j_{\|}|$ should be the same.

For a more delicate view of the plasma characteristics, Figure \ref{fig:fig3}d--\ref{fig:fig3}g plot the time variation of the normalized reduced distribution functions (NR-VDFs) for ions and electrons, respectively. The reduction process along $\bm B$ direction is achieved by double integration of the distribution functions in the $-\bm V \times \bm B$ and $\bm B \times \bm V \times \bm B$ direction. Likewise, the reduction along $\bm B \times \bm V \times \bm B$ is obtained from the double integration of the VDFs in the $\bm B$ and $-\bm V \times \bm B$ directions. The normalization is then completed by dividing the reduced VDFs by its maximum value. First, the NR-VDFs for the ion are examined in Figure \ref{fig:fig3}d and \ref{fig:fig3}e. As shown in Figure \ref{fig:fig3}d, the NR-iVDFs along $\bm B$ are changing dynamically with the broadening and narrowing in the velocity width taking place alternatively. This correspond to the parallel temperature variations shown in Figure \ref{fig:fig3}b. Moreover, beam-like populations drifting at velocity of $\sim$300--400 \vunit (in comparison with the local \Alfven speed $\sim$250 \vunit) are found to appear near $r_2$ and $r_0$. This minor population could contribute up to 10 \% of the major NR-iVDFs centered at $V_B=0$ and thus lead to an asymmetry of the NR-iVDFs with respect to $V_\|=0$. As revealed in Figure \ref{fig:fig3}e, the NR-iVDFs along $\bm B \times \bm V \times \bm B$ are centered at 200 \vunit, which correspond to the $\bm E \times \bm B$ convection motion. In addition, a slight broadening is visible near $r_1$ and it agrees with the perpendicular temperature increase in Figure \ref{fig:fig3}b. Next, the electron observations are presented in Figure \ref{fig:fig3}f and \ref{fig:fig3}g. As seen in Figure \ref{fig:fig3}f, the NR-eVDFs along $\bm B$ exhibit significant variations in the velocity width, but their symmetries with $V_\|=0$ are maintained. In addition, the broadening near $r_1$ and narrowing near $r_2$ and $r_0$ is in reverse trend as compared with NR-iVDFs. The NR-eVDFs along $\bm B \times \bm V \times \bm B$ in Figure \ref{fig:fig3}g stay mostly stable, which explains the generally constant behavior of the perpendicular electron temperature. Last, we note that besides the results shown in the above two directions, the NR-VDFs along $-\bm V \times \bm B$ for both ions and electrons remain nearly unchanging during the whole interval, which can be partly shown as below.

To highlight the distinctive VDFs within the vortex, we show 5 columns of the VDFs (5 snapshots from $t_2$ to $t_6$) in different planes constructed by local $\bm B$, $\bm B \times \bm V \times \bm B$, and $-\bm V \times \bm B$ directions. Four types of VDFs can be identified as below: (1) The iVDFs display beam-like structures on the positive $V_\|$ side of the distribution at $t_2$ (Figure \ref{fig:fig3} h1), $t_4$ (Figure \ref{fig:fig3} j1). These beams seem to partially merge with the major population at $t_6$ (Figure \ref{fig:fig3} l1). (2) The perpendicular broadenings of the iVDFs in the $\bm B \times \bm V \times \bm B$ direction are evident at $t_3$ (Figure \ref{fig:fig3} i1, i2) and $t_5$ (Figure \ref{fig:fig3} k1, k2). (3) The eVDFs display clear elongation on both the positive and negative side of the parallel direction at $t_3$ (Figure \ref{fig:fig3} n1) and $t_5$ (Figure \ref{fig:fig3} p1), with the appearance of bi-directional beam-like structures at $v$ $\sim$ 6000 \vunit. (4) The eVDFs are generally isotropic at $t_2$, $t_4$ (Figure \ref{fig:fig3} m1, o1, m2, o2), and are slightly anisotropic at $t_6$ (Figure \ref{fig:fig3} q1, q2). If examine more carefully at these times, it can be found that the contours of eVDFs in the anti-parallel direction are closer than the ones in the positive direction (see the dense contours at $v$ $\sim$ --4000 \vunit in Figure \ref{fig:fig3} m1, o1, p1). This has led to a net negative electron drift velocity much larger than the ion parallel velocity, and thus are responsible for the parallel current near $r_0$ and $r_2$. In particular, the snapshots of eVDFs presented here are reminiscent of the electron distributions observed in a fast solar wind event by \citet{2017ApJ...849...49P}, where the authors have reported an isotropic distribution in the vortex center and an increased phase space density at the vortex boundary. However, the anisotropic characteristics are somehow different since the ``strahl'' populations from the background solar wind, which affect the distribution in the parallel direction, are always present in \citet{2017ApJ...849...49P}.

\section{Energy conversion channels associated with the Alfve\'n vortex} \label{sec:energy}
The deformation of the particle distributions, as seen in both gyrotropic and non-gyrotropic temperature anisotropy, have been reported in space observations and simulations of plasma turbulence \citep{1982JGR....87...52M,2015ApJ...800L..31H,2011ApJ...739...54V,2012PhRvL.108d5001S,2013ApJ...762...99P,2016AIPC.1720d0003F}. Despite the active debate concerning the kinetic scale nature of the turbulence (i.e. waves and/or structures in \citeauthor{2018arXiv180605741G} \citeyear{2018arXiv180605741G}), two types of mechanism have been widely invoked to explain such phenomenon. One is wave-particle interactions, such as cyclotron and Landau resonances with kinetic \Alfven and slow-mode waves \citep{2015ApJ...800L..31H,2015ApJ...813L..30H}. In a more recent paper, modulations of the ion and electron pitch angle in the presence of large-amplitude electromagnetic waves are also found \citep{2018ApJ...867...58Z}. The other is dissipation near coherent structures \citep{2012PhRvL.108z1102O,2012PhRvL.109s5001W,2016PhRvL.116n5001P}. This mechanism is related to gradients in the magnetic/velocity field, or more specifically, the work done by the pressure-strain interaction $-(\bm P \cdot \bm \nabla) \cdot \bm u_F$, which can be decomposed into the isotropic compression term $-p\theta$ and traceless pressure-strain interaction term $-\mathit{\Pi}_{ij}D_{ij}$. Here $\bm P$ is the pressure tensor, $\bm u_F$ is the bulk flow velocity, $p=1/3 P_{ii}$ is the scalar pressure, $\theta=S_{ii}$ is the trace of the strain rate tensor, $\mathit{\Pi_{ij}}=P_{ij}-1/3P_{ii}\delta_{ij}$ is the traceless pressure tensor, $D_{ij}=S_{ij}-1/3S_{ii}\delta_{ij}$ is the traceless strain rate tensor, $S_{ij}=1/2(\partial_iu_j+\partial_ju_i)$ is the symmetric strain rate tensor, and $\delta_{ij}$ is the Kronecker delta. As shown in fully kinetic simulations, the pressure work could trigger individual energy conversion channels (for both ions and electrons) between fluid energy and random thermal energy \citep{2017PhRvE..95f1201Y,2017PhPl...24g2306Y}. This idea has been tested in a few current layers (see the MMS observation of electron energy conversion channel in \citeauthor{2018ApJ...862...32C} \citeyear{2018ApJ...862...32C}). More importantly, theoretical models have proved recently that, the momentum anisotropy contained in a sheared flow could lead to proton pressure anisotropy from an initial isotropic state \citep{2016PhRvE..93e3203D,2018MNRAS.475..181D}. To be more precise, the evolution of gyrotropic and non-gyrotropic anisotropies are driven by $-\mathit{\Pi}_{ij}D_{ij}$ term, while the compression term seems not contributing \citep{2018MNRAS.475..181D}.

\begin{figure*}
\figurenum{4}
\includegraphics [clip,trim=0cm 6cm 0cm 0cm, width=1.00\textwidth]{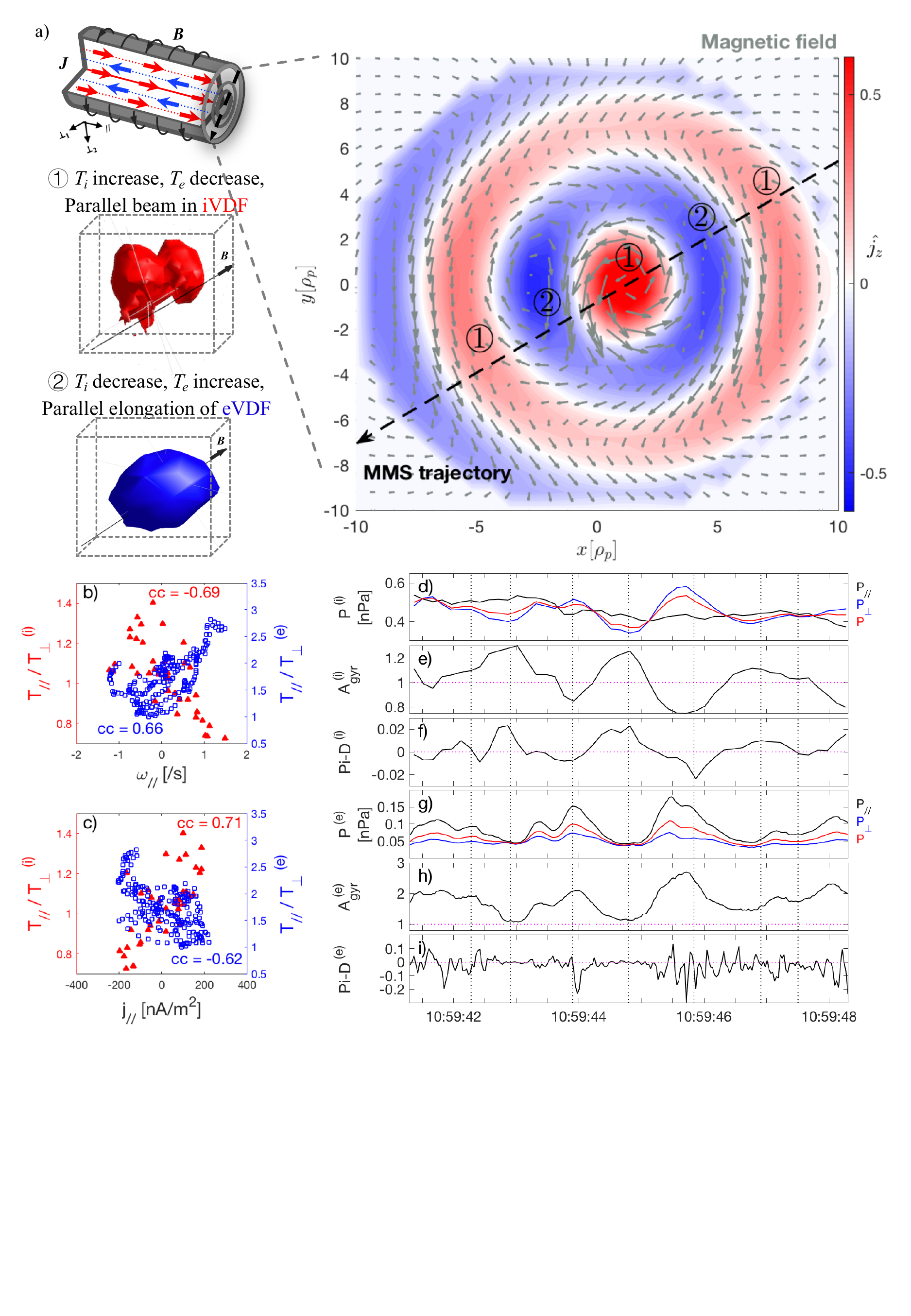}
\caption{
Illustration of the kinetic signatures within the \Alfven vortex. a) The geometry, electromagnetic field, velocity field of the vortex, together with the 3D view of two non-Maxwellian velocity distribution functions. b) Scatter plot of parallel vorticity versus ion and electron temperature anisotropy. c) Scatter plot of parallel current density versus ion and electrons temperature anisotropy. d) Ion parallel, perpendicular and total pressure. e) Ion gyrotropic pressure anisotropy $A_{gyr}^{(i)}=P_{\|}/P_{\bot}^{(i)}$. f) Ion pressure-strain interaction term $P_i-D^{(i)}$. g)-i) Same as d)-f) but for electrons.
\label{fig:fig4}}
\end{figure*}

To find a possible interpretation for the observed pressure anisotropies, we have thus investigated the ion and electron $-\mathit{\Pi}_{ij}D_{ij}$ terms within the vortex (denoted as $P_i-D^{(i)}$ and $P_i-D^{(e)}$, respectively). The stress tensor is obtained using the curlometer technique, which also provide gradient estimation for the velocity field \citep{2002JGRA..107.1384D}. Figure \ref{fig:fig4}d--\ref{fig:fig4}i plot the pressure components, pressure anisotropy $A_{gyr}=P_{\|}/P_{\bot}$, and $P_i-D$ for ions and electrons, respectively. It can be seen that the ion pressure shows variations in both parallel and perpendicular directions, whereas the electron pressure is mostly varying in the parallel direction. Despite the larger pressures of ions compared with electrons, $P_i-D^{(e)}$ is larger than $P_i-D^{(i)}$ for nearly one order of magnitude. In addition, ${P_i-D^{(e)}}$ exhibits multiscale variations where it not only contains variations of similar scale and amplitude as its ion counterpart, but also comprises many sub-ion scale structures with higher amplitude. These results indicate a stronger and more complex pressure-strain interaction accompanied by the electron flow-induced strains. Furthermore, If we compare the trend between $-\mathit{\Pi}_{ij}D_{ij}$ and $A_{gyr}$, it can be found that $P_i-D^{(i)}$ changes almost simultaneously in phase with $A_{gyr}^{(i)}$ (Figure \ref{fig:fig4}e--\ref{fig:fig4}f), whereas $P_i-D^{(e)}$ changes in anti-phase with $A_{gyr}^{(e)}$(Figure \ref{fig:fig4}h--\ref{fig:fig4}i). Although the causal relation is not necessarily implied, the correlations may reflect an inherent link between the time-series of the work done by the pressure-stress interaction and the pressure anisotropy. Specifically, the correlation between ${P_i-D^{(i)}}$ and $A_{gyr}^{(i)}$ found here seems to agree with the scenario proposed by \citet{2018MNRAS.475..181D}, but the correlation for $P_i-D^{(e)}$ and $A_{gyr}^{(e)}$ still requires future theoretical explorations.

\section{Conclusions and discussion}
\label{sec:conclusion}
In conclusion, we have analysed for the first time, plasma properties within an \Alfven vortex embedded in the Earth's turbulent magnetosheath. This in-situ observation is made possible, attributing to the high temporal particle measurements from the MMS mission. As illustrated in Figure \ref{fig:fig4}a, the \Alfven vortex has a radius around 10 proton gyroradius and is identified as a two-dimensional quasi-monopole type. The magnetic field and velocity field are rotating along the vortex axis, mostly in the azimuthal directions. Within the vortex, both ions and electrons exhibit distinctive characteristics which separate from the ambient plasma:
\begin{enumerate}
\item The ion temperature displays variations within the vortex, which are correlated with the parallel current density: It reaches local maximum in the vortex centre, then goes down and arrives to its local minimum within the inversed current, finally it increases again in the vortex edge. Electrons behave in an opposite way as compared with ions, where its temperature variations are correlated with the parallel vorticity: It reaches local maximum near the vortex edge and the inversed current, while having the local minimum in the vortex center.
\item Ions are parallel anisotropic $T_{\|}>T_{\bot}$ within the $j_{\|}>0$ regions and perpendicular anisotropic $T_{\bot}>T_{\|}$ within the $j_{\|}<0$ regions. Electrons, on the contrary, are isotropic within the $j_{\|}>0$ regions and parallel anisotropic $T_{\|}>T_{\bot}$ within the $j_{\|}<0$ regions. The temperature anisotropies for both components are correlated well with parallel current density/vorticity (see Figure \ref{fig:fig4}b and \ref{fig:fig4}c). The strongest ion anisotropy (i.e., $T_{\|}/T_{\bot}$ = 0.7 and 1.4) occurs in the strong shear regions where the gradient of the field reaches maxima ($j_{\|}$ $\sim$ $\pm$ 200 \junit), but the strongest electron anisotropy ($T_{\|}/T_{\bot}$=2.5) only occurs at local minimum of $j_{\|}$ $\sim$ --200 \junit. The isotropic states of ions happen at crossings of the zero current (zero of the Bessel functions in the vortex solutions), while the isotropy of electrons happens at local maximum of $j_{\|}$ $\sim$ 200 \junit.
\item Deformations of the VDFs, featuring elongations along or across the magnetic field, are modulated by $|\nabla v_{\bot}|$ and $|\nabla b_{\bot}|$. In particular, ion beams with positive parallel drifting speed (marked as $\textcircled{\small{1}}$ in Figure \ref{fig:fig4}a), together with bi-directional parallel electron beams (marked as $\textcircled{\small{2}}$ in Figure \ref{fig:fig4}a) are found within the vortex.
\end{enumerate}

These results provide observational evidence of local kinetic processes within the \Alfven vortex, which may help to understand the intermittent heating and energy transfer processes within the coherent structures. In addition, the non-thermal temperature anisotropies and the deformations of ion and electron VDFs might introduce instabilities to the system (i.e. cyclotron type when $T_{\bot}>T_{\|}$ or firehose type when $T_{\|}/T_{\bot}$), which may in turn, affect the small-scale turbulence cascade. Though only one event of a well-defined \Alfven vortex is presented in this paper, a statistical study is being performed to further investigate the relation between \Alfven vortices and plasma kinetic effects in different scenarios.

One limitation of the current work is the interpretation of the fluctuations as described by the classical shear \Alfven vortex model, which is solution of the Kadomtsev-Pogutse-Strauss system of the reduced incompressible MHD equations \citep{1992swpa.book.....P}. This model only considers the nonlinear effects of the shear \Alfven waves, while the compressible effects have been neglected. Nevertheless, we note that the \Alfven vortex observed here, being similar with \Alfven vortices in the slow solar wind \citep{2016ApJ...826..196P}, is in fact compressible. To describe the compressive coherent magnetic vortices in high-beta plasma, \citeauthor{2017arXiv170502913J} (submitted to APJ, arxiv \citeyear{2017arXiv170502913J}) has developed a new model. By omitting the heat flux and thus considering the equations of state, the normalized density fluctuations $\delta \hat{n}=\delta n/n_0$ and compressible magnetic field fluctuations $\delta \hat{B}_{\|}=\delta B_{\|}/B_0$ are solved via the generalized pressure balance condition. For the solutions at scales larger than ion Larmor radius, $\delta \hat{n}$ and $\delta \hat{B}_{\|}$ are localised within the vortex core. Their relative ratio, also known as plasma compressibility \citep{1986JPlPh..35..431G}, is expressed as
\begin{equation}
\begin{aligned}
\label{equ:equ5}
C_{plasmas}&= \frac{\delta \hat{n}}{\delta \hat{B}_{\|}}\\
&=\frac{\frac{\beta_{i\bot}}{4}+\frac{V_A^2}{u_z^2}(1-\frac{\beta_{e\|}-\beta_{e\bot}}{2})}
{\frac{\beta_{i\bot}}{4}-\frac{V_A^2}{u_z^2}
\frac{\gamma_{i\bot}\beta_{i\bot}+\gamma_{e\bot}\beta_{e\bot}}{2}(1-\frac{\beta_{e\|}-\beta_{e\bot}}{2})}
\end{aligned}
\end{equation}

Where $u_z$ is the vortex speed along the magnetic field, $\gamma_{i\bot}$ and $\gamma_{e\bot}$ are the polytropic indices for the ions and electrons, satisfying $T_{s,\bot} \propto n_s^{\gamma_{\bot}-1}$. In comparison with this model, we find that: (1) $\delta B_{\|}$ are localized within the vortex, while $\delta B_{\bot}$ ``leaks out'' from the core to larger distances (Figure \ref{fig:fig2}a). The localization of $\delta B_{\|}$ is associated with the 3 seconds scale of the mean field, while the variations in $B_{abs}$ (Figure \ref{fig:fig1}) cover a wider range. (2) The observed $C_{plasmas}$ (for the time scales from 0.3 to 3 s) could reach $\sim$ 2 in the vortex core ($\delta n/n_0$ $\sim$ 0.32, $\delta B_{\|}/B_0$ $\sim$ 0.15), while the theoretical mean value estimated from equation (\ref{equ:equ5}) is around 4, if we take $u_z/V_A=u_\bot /\tan(\theta_{vortex})/V_A$ $\sim$ 1.7, $\beta_{i\bot}$ =0.9, $\beta_{e\|}$ =0.18, $\beta_{e\bot}$ = 0.1, and use $\gamma_{i\bot}$ =0.58 $\pm$ 0.13, $\gamma_{e\bot}$ =1.1 $\pm$ 0.04 as fitted from the density and temperature measurement. Hence, the observed compressible features qualitatively agree with the theory of \citet{2017arXiv170502913J}. It should be noted that $C_{plasma}$ appears to be a sensitive function of $\theta_{vortex}$. For example, if $\theta_{vortex}$ is larger than 8$^\circ$, $\delta B_{\|}/B_0$ may touch zero and $C_{plasma}$ becomes infinite in our case, while if $\theta_{vortex}$ is zero, $u_z$ is infinite and $C_{plasma}$ is 1. We also remark that the double polytropic equations, although lacking the kinetic features, may serve as a specific description for the thermal anisotropies of coherent structure (e.g., Interpretation of magnetic holes in \citeauthor{2018ApJ...864...35Z} \citeyear{2018ApJ...864...35Z}). Future attempts on basis of polytropic laws need to be made to address the compressibility and thermal dynamics within the \Alfven vortex.

\acknowledgements{
We greatly appreciate the MMS development and operations teams, as well as the instrument PIs for data access and support. We thank the reviewer for valuable suggestions on the interpretation and compressibility of the vortex. We thank Jian Yang for help with MMS data analysis. This work was supported by the Marie Sk\l odowska-Curie grant No. 665593 from the European Union's Horizon 2020 research and innovation programme. Work by D. Perrone was supported by STFC grant ST/N000692/1. Work by O. Alexandrova was supported by the French Centre National d'Etude Spatiales (CNES). The data are publicly available from https://lasp.colorado.edu/mms/sdc/public/.}


\bibliographystyle{aasjournal}
\bibliography{classify}





\end{document}